\documentclass[twocolumn]{book}
\usepackage[dvips]{graphicx,color}
\usepackage{makeidx,universe}
\usepackage{amssymb}
\usepackage{hyperref}

\makeauthorindex

\BookTitle{Proceedings of the XXIX PHYSICS IN COLLISION}

\CopyRight{\copyright 2009 by Universal Academy Press, Inc.}

\begin{document} 

\pagenumbering{arabic}

\chapter{%
{\LARGE \sf
High $E_T$ Jet Production} \\
{\normalsize \bf 
Kenichi Hatakeyama } \\
{\small \it \vspace{-.5\baselineskip}
Baylor University, 
One Bear Place, Waco, TX, 76798, USA 
}
}


\AuthorContents{K.\ Hatakeyama}

\AuthorIndex{Hatakeyama}{H.} 

\baselineskip=10pt 
\parindent=10pt    

\section*{Abstract} 

A review is presented on studies of high $E_T$ jet production
and production of photon, $W$ and $Z$ associated with jets
from HERA and Tevatron experiments.
Such studies have been used to examine the Standard Model
(SM) in the area of the strong interaction,
Quantum Chromodynamics,
at highest energies currently attainable in collider
experiments,
to extract values of the coupling of the strong interaction,
to determine the parton distribution functions in the proton,
and to provide constraints on SM processes that constitute
background to the Higgs boson and new physics searches.
Some of them are also directly sensitive to the presence
of physics beyond the SM.
Future prospects for results from the LHC experiments are discussed.

\section{Introduction} 
\label{sec:intro}

In this paper, we will discuss studies of high $E_T$ jet
production and production of photon, $W$ and $Z$ associated
with jets from
experiments at the DESY $ep$ collider, HERA, Hamburg, Germany
and experiments at the Fermilab $p\bar p$ collider, Tevatron,
Illinois, USA.
HERA collides electrons or positrons with protons at
$\sqrt{s}\sim 319$ GeV 
and it delivered
approximately 0.5 fb$^{-1}$ of data per experiment
from 1999 until the
end of its operation in 2007 (HERA II).
The ZEUS and H1 are two general purpose experiments
at HERA.
The Tevatron collides protons and antiprotons
at $\sqrt{s}=1.96$ TeV, and since 2001 until now (Run II),
it delivered about 7 fb${}^{-1}$ of data per experiment\footnote{The results
presented here use data up to 2.5 fb${}^{-1}$.}.
Two experiments are operating on the Tevatron; CDF and D0.
The results from these experiments are reviewed in this paper.

\begin{figure}[th]\centering\leavevmode
\centerline{
  \includegraphics[width=0.30\columnwidth]
  {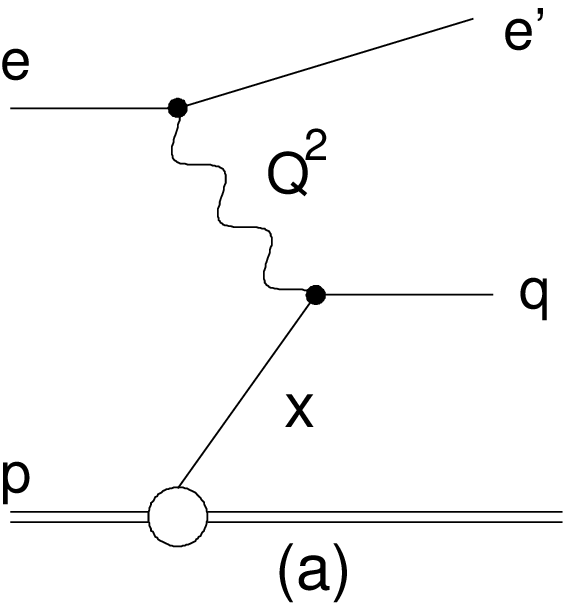}
  \includegraphics[width=0.30\columnwidth]
  {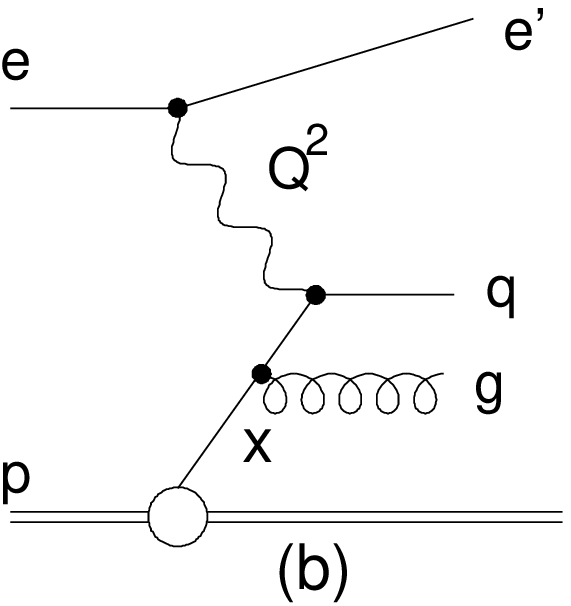}
  \includegraphics[width=0.30\columnwidth]
  {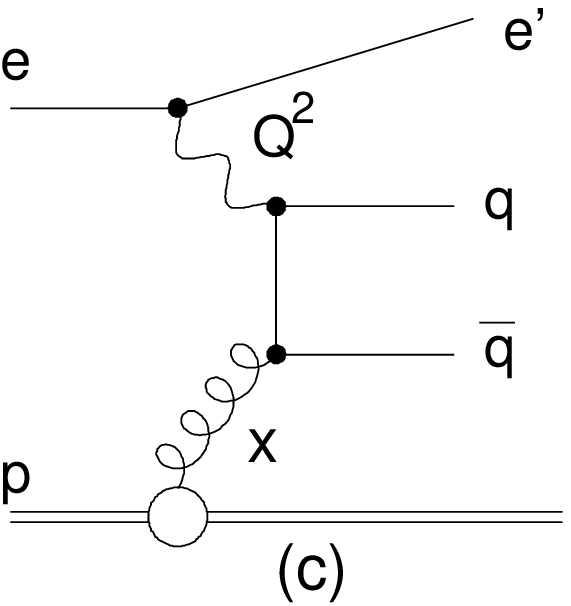}
}
\caption{Deep inelastic $ep$ scattering processes:
{\em (a)} Born contribution, {\em (b)} example of the QCD Compton
and {\em (c)} boson-gluon fusion processes~\cite{Aaron:2009vs}.}
\label{fig:NCDIS-jets}
\end{figure}

%
%
%
Jet production diagrams in neutral current deep inelastic
scattering (NC DIS) in $ep$ collisions are shown in
Fig.~\ref{fig:NCDIS-jets}.
At the $\alpha\alpha_s^0$ order there is the Born process,
and at the $\alpha\alpha_s^1$ order there are contributions
from the QCD Compton process and
boson-gluon fusion process.
In the Breit frame
in which the virtual photon and the proton collide head-on,
the jet from the Born process does not have any transverse momentum,
thus the lowest order non-trivial contributions to high $p_T$
jets come from the QCD Compton and boson-gluon fusion
processes.
The jet production cross section in NC DIS can be expressed
in QCD as:
\begin{equation}
d\sigma_{jet}=
\sum_{a=g,q,\bar q} f_{a/p}(x,\mu_F^2) 
\hat\sigma_{a}(x,\alpha_s(\mu_R^2))dx
\end{equation}
where $f_{a/p}$ is the parton distribution function (PDF)
of the proton, $\hat\sigma_a$ is the partonic subprocess
cross section,
and $\mu_R$ and $\mu_F$ are the
renormalization and factorization scales, respectively.
By measuring jet production cross sections
and comparing them with perturbative QCD predictions,
we test QCD factorization and
universality of the PDFs and $\alpha_s$.
And measurements are also used to extract
PDFs of the proton and $\alpha_s$.
In addition to NC DIS, in this paper results on jet production
in photoproduction are discussed.
In NC DIS, $Q^2>>0$, while in photoproduction, 
$Q^2\sim0$, where $Q^2$ is the negative momentum transfer squared.
%

\begin{figure}[th]\centering\leavevmode
\centerline{
  \includegraphics[width=0.70\columnwidth,bb=100 160 560 440,clip=]
  {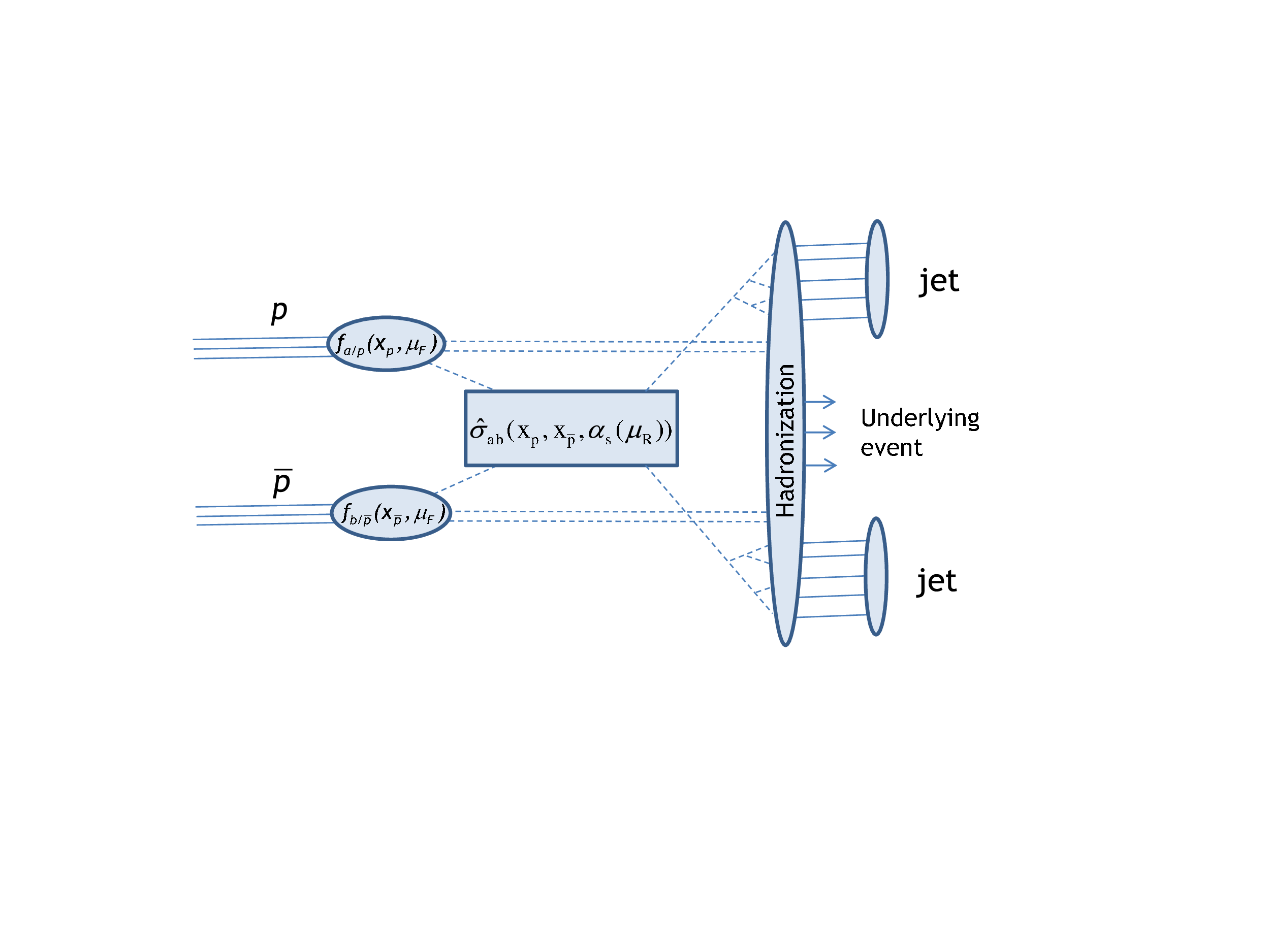}
}
\caption{Schematic drawing for
  jet production in proton-antiproton collisions.}
\label{fig:jets_ppbar}
\end{figure}

%
%
The schematic drawing of jet production in $p\bar p$
collisions is shown in Fig.~\ref{fig:jets_ppbar}, and its
production cross section can be expressed
in QCD as:
\begin{eqnarray}
d\sigma_{jet}&=&
\sum_{a=g,q,\bar q}\sum_{b=g,q,\bar q}
f_{a/p}(x_p,\mu_F^2)
f_{b/\bar p}(x_{\bar p},\mu_F^2)\nonumber\\
&&~~~~~
\hat\sigma_{a,b}(x_p,x_{\bar p},\alpha_s(\mu_R^2))dx_{p}dx_{\bar p}.
\end{eqnarray}
The experimental measurements of jet cross section
at the Tevatron also provide stringent test of QCD predictions,
information on $\alpha_s$,
and also powerful constraints on proton PDFs.
In addition, some measurements can be used to search
for physics beyond the SM (BSM) such as exotic heavy
particles decaying into two jets, quark compositeness
and large extra dimensions.
In hadron-hadron collisions,
the underlying event, everything except the hard scattering
contributions, makes jet measurement complicated as the
underlying event overlaps with jets, and
good understanding of the underlying event plays an important role
in jet physics.

In Sec.~\ref{sec:incjet_dijets},
measurements on inclusive jet and dijet production at HERA
and Tevatron are reviewed.
Soft QCD physics studies from HERA and Tevatron,
such as measurements
of the underlying event, inclusive $p\bar p$ collisions and
double parton interactions are discussed later in
Sec.~\ref{sec:ue}.

In Secs.~\ref{sec:photon_jets} and \ref{sec:wz_jets},
studies on photon, $W$, and $Z$ production in
association with jets from HERA and Tevatron are discussed.
The photon, in comparison with jets, is the direct probe
of hard scattering as it is not affected by hadronization.
The measurements on photon production in association with jets
are also important tests of QCD and have a good potential
to constrain proton PDFs.

$W/Z$+jets events constitute important event
samples for physics at the Tevatron and LHC.
Many top quark measurements, searches
for the Higgs boson, Supersymmetry (SUSY) and other
BSM phenomena begin with $W/Z$+jets event samples.
Good understanding of $W/Z$+jets production from QCD
is critical for such studies.
The perturbative QCD calculations and many Monte Carlo (MC) tools are
available for $W/Z$+jets production and they need to be validated
by experimental measurements. These are discussed in
Sec.~\ref{sec:wz_jets}.

Future prospects for results from the LHC experiments in soft
QCD and on high $E_T$ jets are discussed in Sec.~\ref{sec:lhc}.

\subsection{Jet Algorithms and Jet Corrections}

%
%
Jets are clusters of particles; and thus, for quantitative
studies, they need to be defined by an algorithm.
Jet algorithms should work in both experimental environment
and theoretical calculations, and allow their comparisons with
minimal ambiguities.
To satisfy this need, a variety of jet algorithms have been
proposed~\cite{Salam:2009jx,Ellis:2007ib,Volobouev:2009rv},
and the properties of those algorithms have been studied in detail.
%
%
%
%

All the measurements from HERA presented in this paper
use the inclusive $k_T$ algorithm.
The $k_T$ algorithm has an advantage over most of seeded cone algorithms
in that it is infrared safe at all order of perturbative QCD calculations.
In the Tevatron experiments, however, seeded cone algorithms such as the
midpoint algorithm have been widely used.
This is mainly due to the simplicity of cone algorithms in constructing
corrections for the underlying event and other experimental effects.

Jets measured by the detectors are affected by various
instrumental and physics effects.
These effects need to be taken into account before
we obtain any physics results and conclusions.
In experimental measurements, jets measured by the
detectors (mostly by calorimeters) are corrected for
any instrumental effects and corrected back
to the particle level, {\it i.e.} jets of stable particles,
and 
can be compared to the theoretical predictions
independently of detector conditions.
The theoretical predictions are often computed for jets of partons
before hadronization.
In such case, the underlying event and hadronization effects
are estimated and the corrections are applied to obtain
the predictions at the particle level.
More details can be found in {\it e.g.} Ref.~\cite{Ellis:2007ib}.

\section{Inclusive jets and dijets}
\label{sec:incjet_dijets}

\subsection{Results from HERA Experiments}

The ZEUS and H1 experiments have provided precise
measurements on inclusive jet and dijet production (see also
the talk by J.~Ferrando~\cite{JamesFerrando}).
These measurements lead to an improved determination
of gluon density in the proton and they
can be also used to extract the value of $\alpha_s$,
one of the fundamental and most important parameters
in QCD.


%
\begin{figure}[tb]\centering\leavevmode
\centerline{\includegraphics[width=0.75\columnwidth]
  {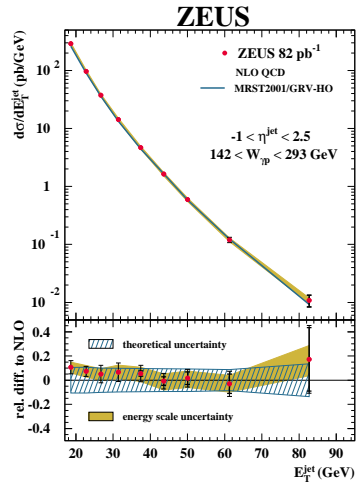}}
\caption{Inclusive jet differential cross section
  as a function of $E_T^{jet}$ in
  photoproduction~\cite{ZEUS-prel-08-008}.}
\label{fig:incjet_pho_zeus}
\end{figure}

The ZEUS experiment recently reanalyzed the published
measurement on inclusive jet production in
photoproduction~\cite{ZEUS-prel-08-008}.
The new analysis uses perturbative QCD calculations up
to the next-to-leading-order (NLO) in $\alpha_s$
with recent parametrizations of the proton PDFs and a
purely theoretical estimation of the uncertainties
coming from higher orders.
The measured inclusive jet differential cross section
is shown in Fig.~\ref{fig:incjet_pho_zeus}.
The $\alpha_s$ value is extracted by comparing it with
perturbative QCD calculations.
The cross section predictions are convolutions
of the proton (and photon in case of photoproduction)
PDFs and the matrix elements.
While the latter depends explicitly on $\alpha_s$,
the former also has an implicit dependence on $\alpha_s$.
The $\alpha_s$ value obtained from this analysis is
$\alpha_s(m_Z)=0.1223\pm0.0001(\mbox{stat.})
^{+0.0023}_{-0.0021}(\mbox{syst.})\pm0.0030(\mbox{th.})$
with a total uncertainty of 3.1\%.
This is one of the most precise measurements of $\alpha_s$
from HERA.


\begin{figure}[tbh]\centering\leavevmode
  \centerline{\includegraphics[width=0.65\columnwidth]
    {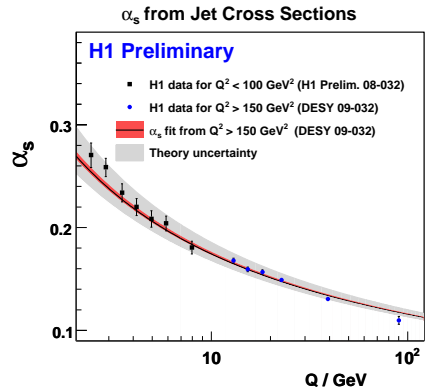}}
  \caption{$\alpha_s$ values obtained from jet measurements
    at low-$Q^2$~\cite{H1-prel-08-032} and
    high-$Q^2$~\cite{Aaron:2009vs} DIS at H1.
  }
  \label{fig:h1_alphas_dis}
\end{figure}

The H1 experiment has recently made measurements on jet
production in NC DIS at low $Q^2$
($5<Q^2<100\mbox{ GeV}^2$)~\cite{H1-prel-08-032}
and high $Q^2$
($150<Q^2<15000\mbox{ GeV}^2$)~\cite{Aaron:2009vs}.
At low $Q^2$, there are a lot of statistics;
however, the extraction of $\alpha_s$ 
suffers from a large theoretical uncertainty
due to $\mu_R$ and $\mu_F$ uncertainties.
The $\alpha_s$ values extracted from 
jet cross sections are shown in
Fig.~\ref{fig:h1_alphas_dis} together with the values
from the high $Q^2$ region~\cite{Aaron:2009vs}.
The $\alpha_s(m_Z)$ from low $Q^2$ is
$\alpha_s(m_Z)=0.1186\pm0.0014(\mbox{exp.})
^{+0.0132}_{-0.0101}(\mbox{th.})\pm0.0021(\mbox{PDF})$
with a total uncertainty of about 10\% dominated by the
scale uncertainty {\it i.e.} by missing higher order
contributions.
The experimental uncertainty largely comes from the jet
energy scale uncertainty which is about $1.5-2$\%.
Higher order QCD calculations will allow extracting
more physics information. 
from DIS data especially at low $Q^2$.

In the recent measurement on inclusive jet
cross sections in high $Q^2$ DIS ($Q^2>125\mbox{ GeV}^2$)
by the ZEUS experiment~\cite{ZEUS-prel-09-006},
the $\alpha_s$ value has been extracted from the differential
cross section $d\sigma/dQ^2$ ($Q^2>500\mbox{ GeV}^2$)
to be
$\alpha_s(m_Z)=0.1192\pm0.0009(\mbox{stat.})
{}^{+0.0035}_{-0.0032}(\mbox{syst.})
{}^{+0.0020}_{-0.0021}(\mbox{th.})$
with a total uncertainty of 3.5\%, largely coming from
the uncertainty in the absolute jet energy scale.

\begin{figure}[tb]\centering\leavevmode
\centerline{
  \includegraphics[angle=270,width=.80\columnwidth]{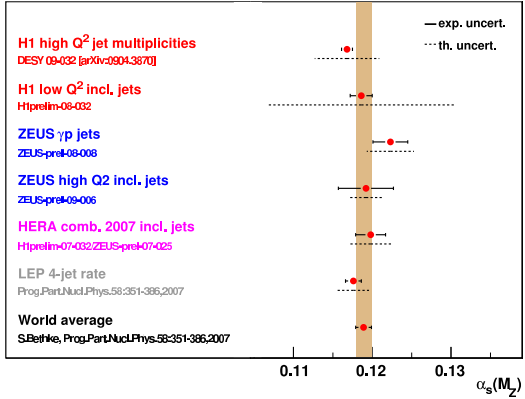}
}
\caption{Summary of recent $\alpha_s(m_Z)$ determinations
  from HERA experiments compared to the
  current world average~\cite{Bethke:2006ac}.}
\label{fig:alphas_hera_summary}
\end{figure}

These $\alpha_s$ measurements are summarized in
Fig.~\ref{fig:alphas_hera_summary} together with
values from the four-jet rate in $e^+e^-$ annihilation 
experiment and the world average.
All $\alpha_s$ values
obtained from many different physics processes
are consistent to each other, indicating a big success of
QCD and universality of $\alpha_s$.
The world average includes results from four-jet rates,
event shape and $Z$ line shape from $e^+e^-$ collisions
at the Large Electron-Positron Collider (LEP) and 
scaling violation and jet measurements from DIS.

\subsection{Results from Tevatron Experiments}

The measurements of
the differential inclusive jet cross section and dijet cross
section at the Tevatron test QCD at the shortest distances
currently attainable in collider experiments,
allow the extraction of $\alpha_s$,
and constrain the parton distribution functions
(PDFs) of the proton, especially gluon densities at high
$x$ ($x\gtrsim0.25$).
They also have unique sensitivities to physics beyond the SM.

The inclusive jet cross section measurements were performed by
CDF using the $k_T$~\cite{CDFIncJetKt} and
midpoint cone~\cite{CDFIncJetCone} algorithms and
using the midpoint algorithm by D0~\cite{DzeroIncJetRPL}.

\begin{figure}[tbh]\centering\leavevmode
\centerline{
  \includegraphics[width=0.70\columnwidth,bb=0 0 567 440,clip=]
  {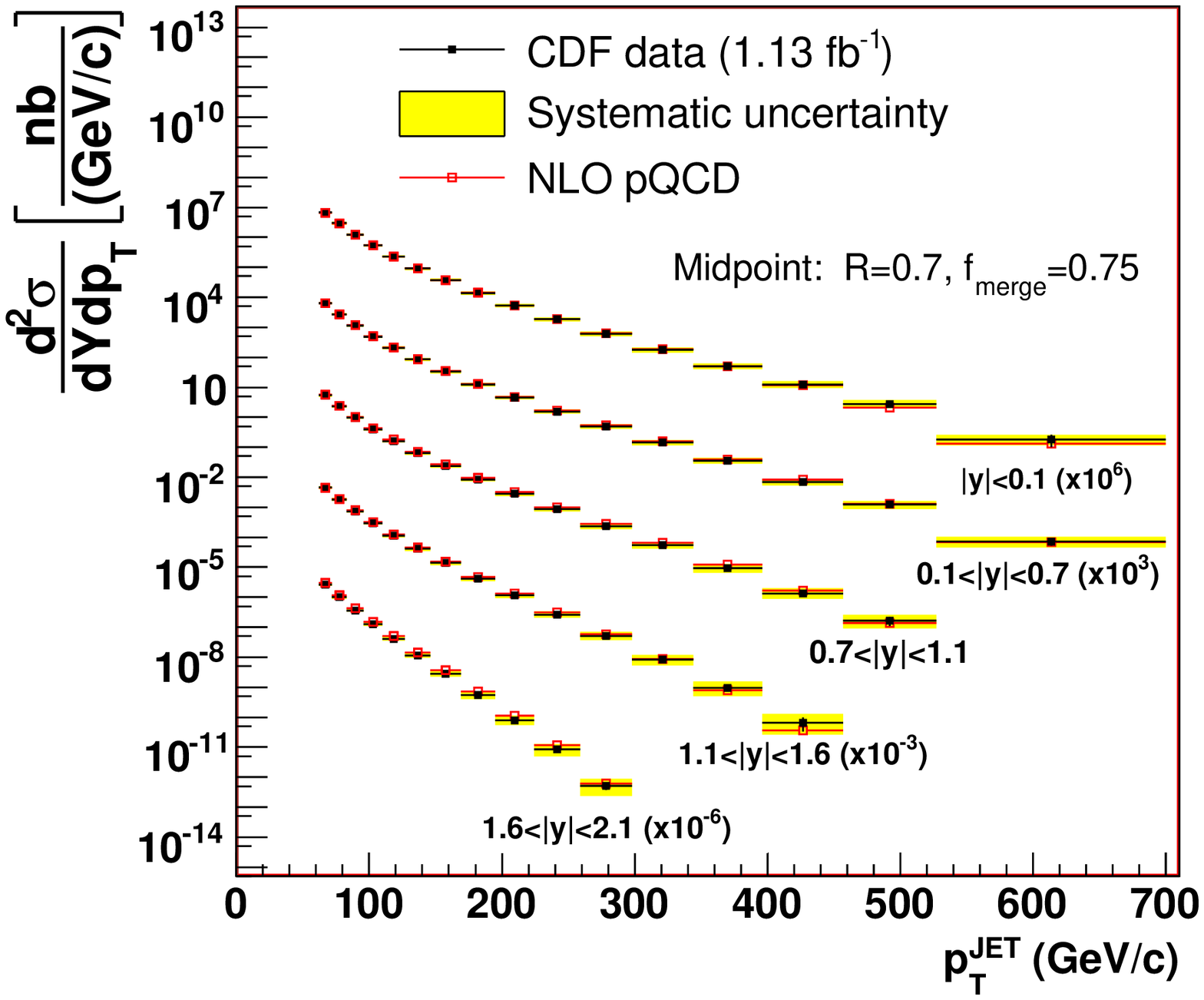}}
\centerline{
  \includegraphics[width=0.90\columnwidth,bb=0 0 567 410,clip=]
  {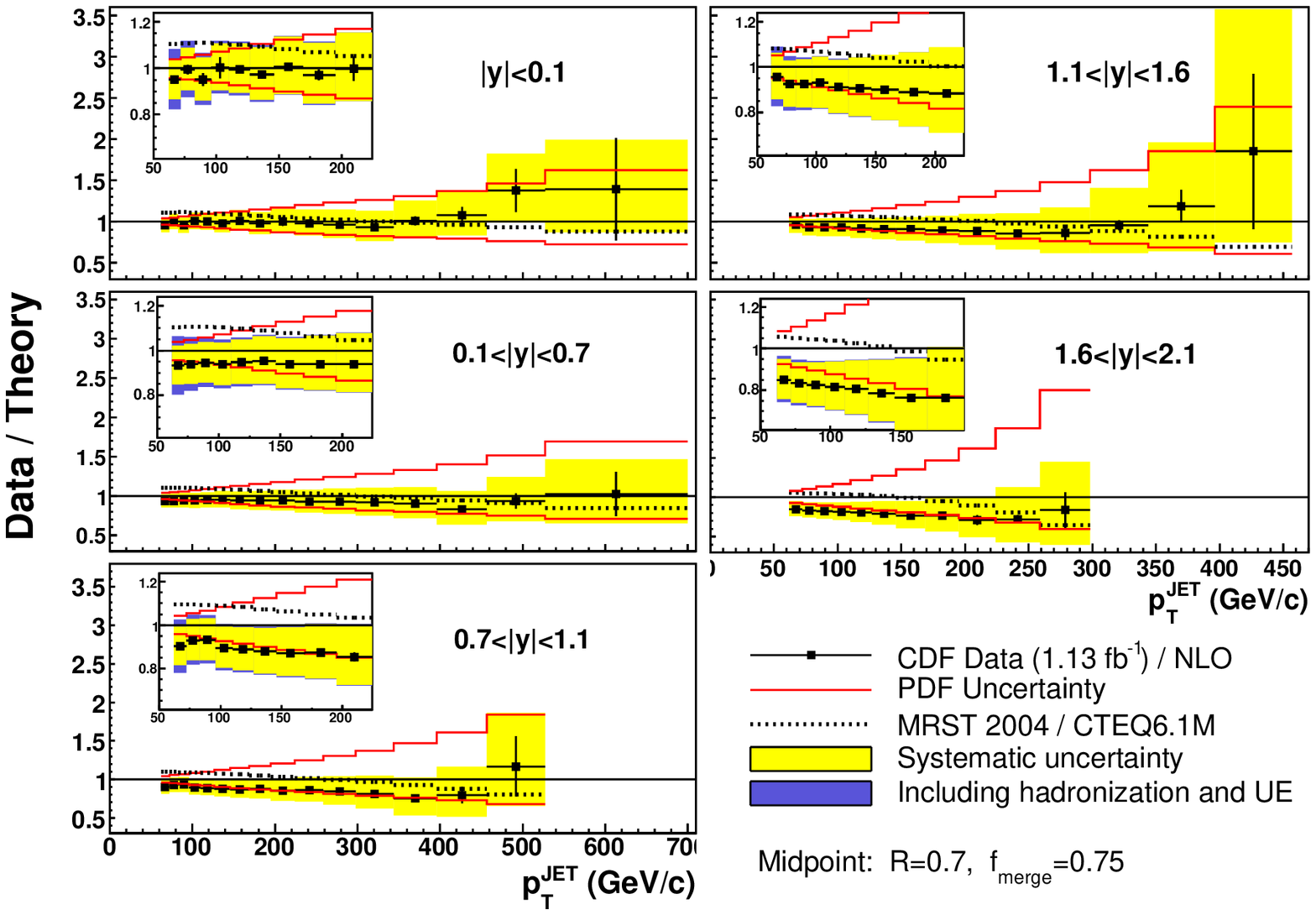}}
\centerline{
  \includegraphics[width=0.90\columnwidth]
  {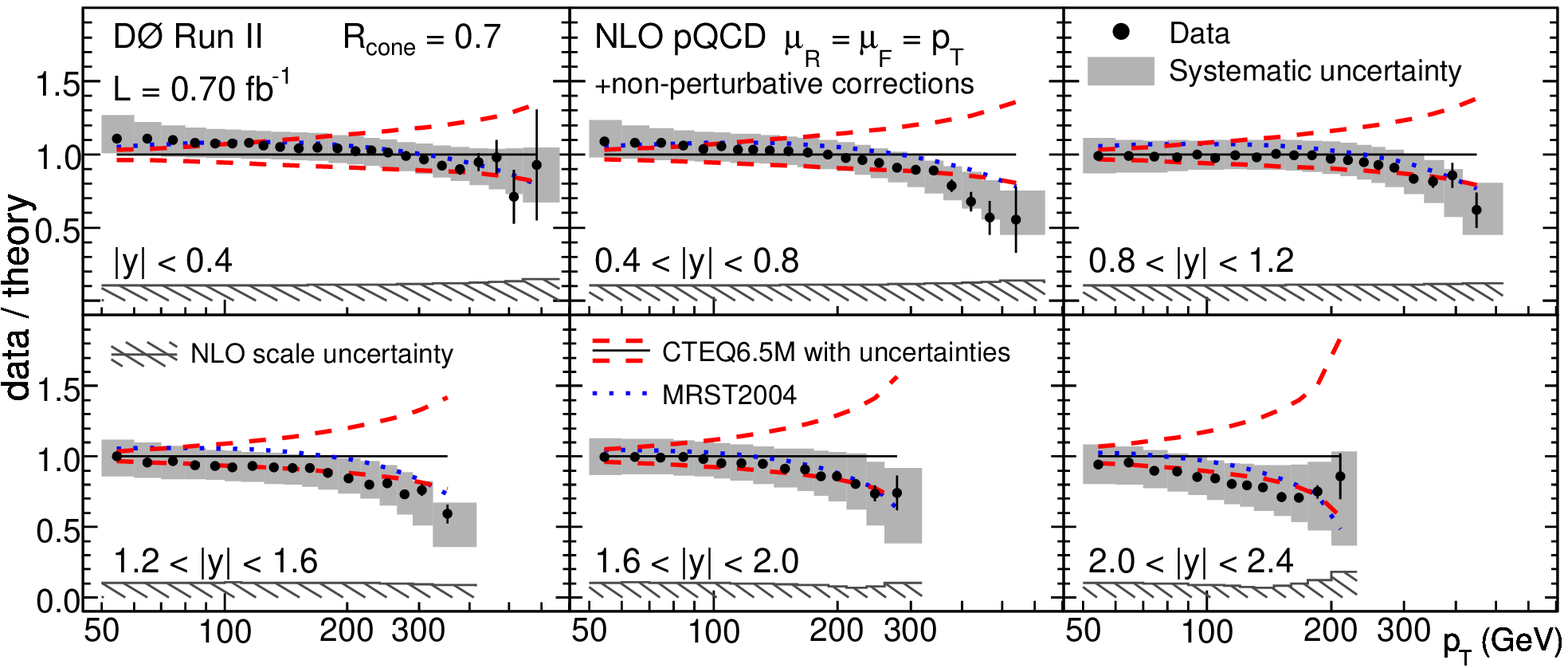}}
\caption{{\em (top)} Measured inclusive jet differential cross sections in
  five rapidity regions by CDF compared to NLO QCD predictions;
  {\em (middle \& bottom)} Ratios of the measured cross sections
  over NLO QCD predictions by CDF and D0.\label{fig:incjet_cdf_d0}}
\end{figure}

The dominant systematic uncertainties in the inclusive jet cross section
measurements arise from the uncertainty in the jet energy scale determination.
The CDF and D0 experiments employ different strategies for the
jet energy scale determination.
The CDF method relies primarily on the measurement of
the response of individual particles~\cite{cdf_JESNIM} in the calorimeter.
Then the calorimeter simulation is tuned based on those measurements,
and the jet fragmentation model is used to determine the jet energy response.
The $\gamma$-jet and $Z$-jet $p_T$ balance is then used for
validation.
In the approach used by D0,
the jet energy scale has been obtained mainly by utilizing the transverse
momentum conservation in $\gamma$+jet events~\cite{dzero_NIM};
in the leading order QCD, the photon and jet are balanced
back-to-back in azimuthal angle $\phi$.
As the photon energy scale is much better known than the
jet energy scale, it is used as a reference for the jet energy scale.
Then the jet energy scale is validated by several cross-checks
in $\gamma$+jet and $Z$+jet events.
With these strategies, 
the CDF and D0 experiments have achieved the jet energy scale
uncertainties of $2-3$\% and $1-2$\%, respectively.
They still lead to $\gtrsim10$\% uncertainty in the differential
cross section as the differential cross section is very rapidly
falling with increasing jet $p_T$, and stay as the dominant
systematic uncertainty in these measurements.

The measured cross sections are compared to the theoretical
predictions in Fig.~\ref{fig:incjet_cdf_d0}.
The measurements are found to be in agreement with NLO QCD
predictions based on then-current PDFs~\cite{CTEQ61,MSTW2004,CTEQ66};
although the data from both CDF and D0 tend to lie somehow on
the lower side of the theoretical predictions.
This trend is consistently seen in the measurements by CDF
using two different jet algorithms and in the measurement by
D0.

Two groups which perform global QCD analyses to determine the
proton PDFs
included these Tevatron inclusive jet cross section
measurements~\cite{CDFIncJetKt,CDFIncJetCone,DzeroIncJetRPL}
in their analyses recently, and the resulting 
PDF sets are referred to as MSTW2008~\cite{MSTW2008} and
CT09~\cite{CT09}.
In these PDF sets,
the inclusion of these measurements lead to somewhat softer
high-$x$ gluons than the ones previously available.
This change is in the direction more consistent with the constraints
from the DIS data.
The better understanding of the proton PDFs due to these measurements
will enhance the discovery potential at the LHC by leading to
better SM background estimation in searches for {\it e.g.}
quark substructure and extra dimensions.

\begin{figure}[thb]
\centerline{\includegraphics[width=0.65\columnwidth]
  {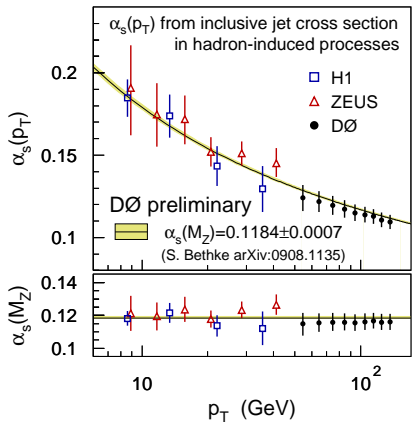}}
\caption{Extracted $\alpha_s(p_T)$ {\em (top)} and 
  $\alpha_s(m_Z)$ {\em (bottom)} values as a function of
  jet $p_T$.}
\label{fig:alphas}
\end{figure}

As HERA experiments could extract $\alpha_s$ from jet cross section
measurements, the Tevatron experiments can also extract $\alpha_s$ from
the jet cross section as well.
CDF performed this analysis using the 1994-95 data~\cite{CDFAlphaS},
and 
D0 recently presented a new $\alpha_s$ determination~\cite{D0AlphaS}
based on the inclusive jet cross section measurement presented
above~\cite{DzeroIncJetRPL}.
In order to avoid the complication arising from the
$\alpha_s$ dependence on PDF determinations,  
only 22 data points out of 110 from this measurement in which
the corresponding parton $x$ of the data point is $x\lesssim0.25$
are used in the extraction of $\alpha_s$, as this measurement
has significant impact on PDFs only at $x\gtrsim0.25$.
This $\alpha_s$ extraction from the Tevatron
extend the HERA results to higher scales 
as shown in Fig.~\ref{fig:alphas}, and they show consistent results.
The extracted $\alpha_s$ value from the D0 analysis is
$\alpha_s(m_Z)=0.1173{}^{+0.0041}_{-0.0049}$ with a
$+3.5-4.2$\% precision which is competitive to results from 
HERA experiments.

\begin{figure}[tbh]
  \centerline{\includegraphics[width=0.60\columnwidth,bb=5 45 480 670,clip=]
    {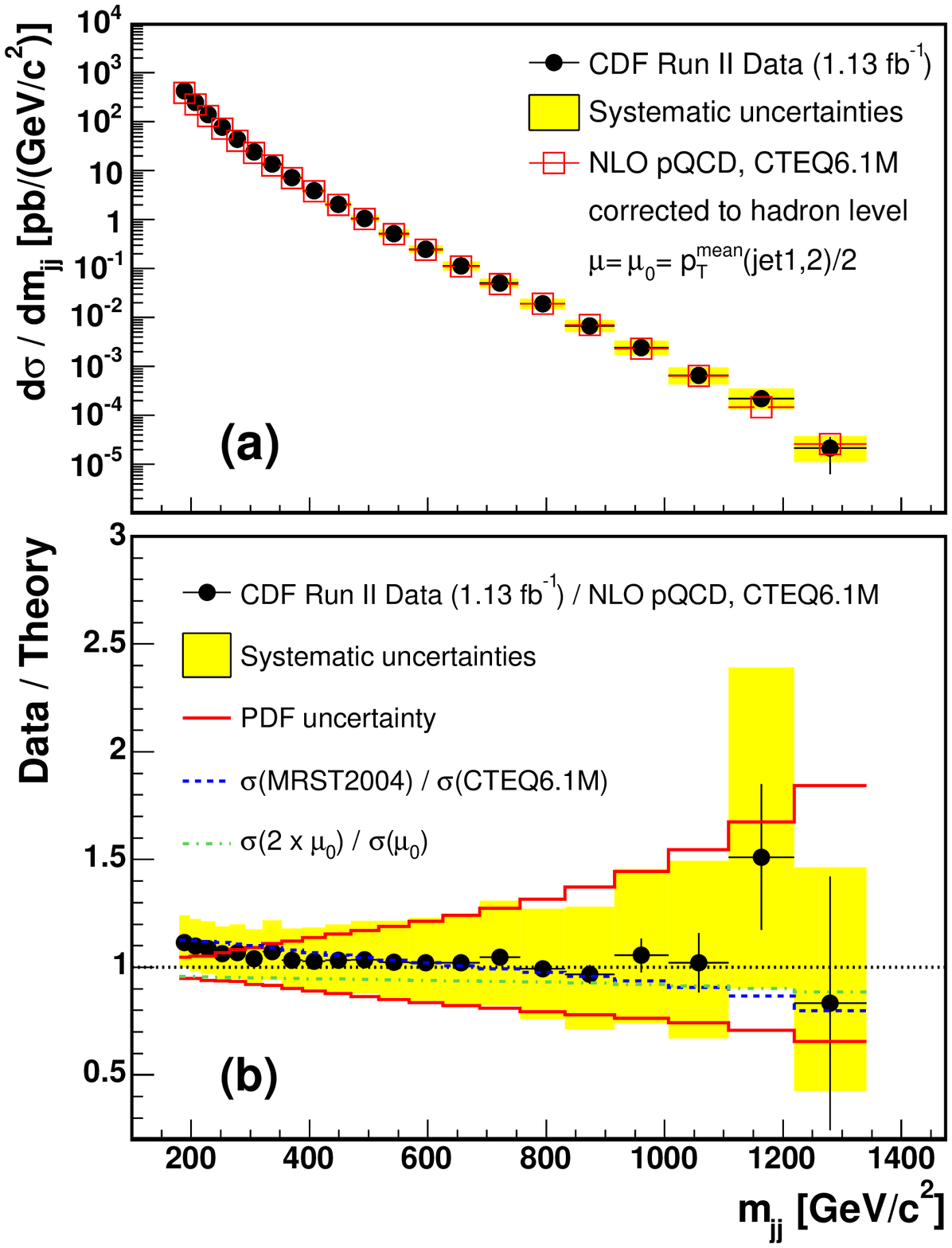}}
   \centerline{\includegraphics[width=0.60\columnwidth,bb=0 0 550 624,clip=]
     {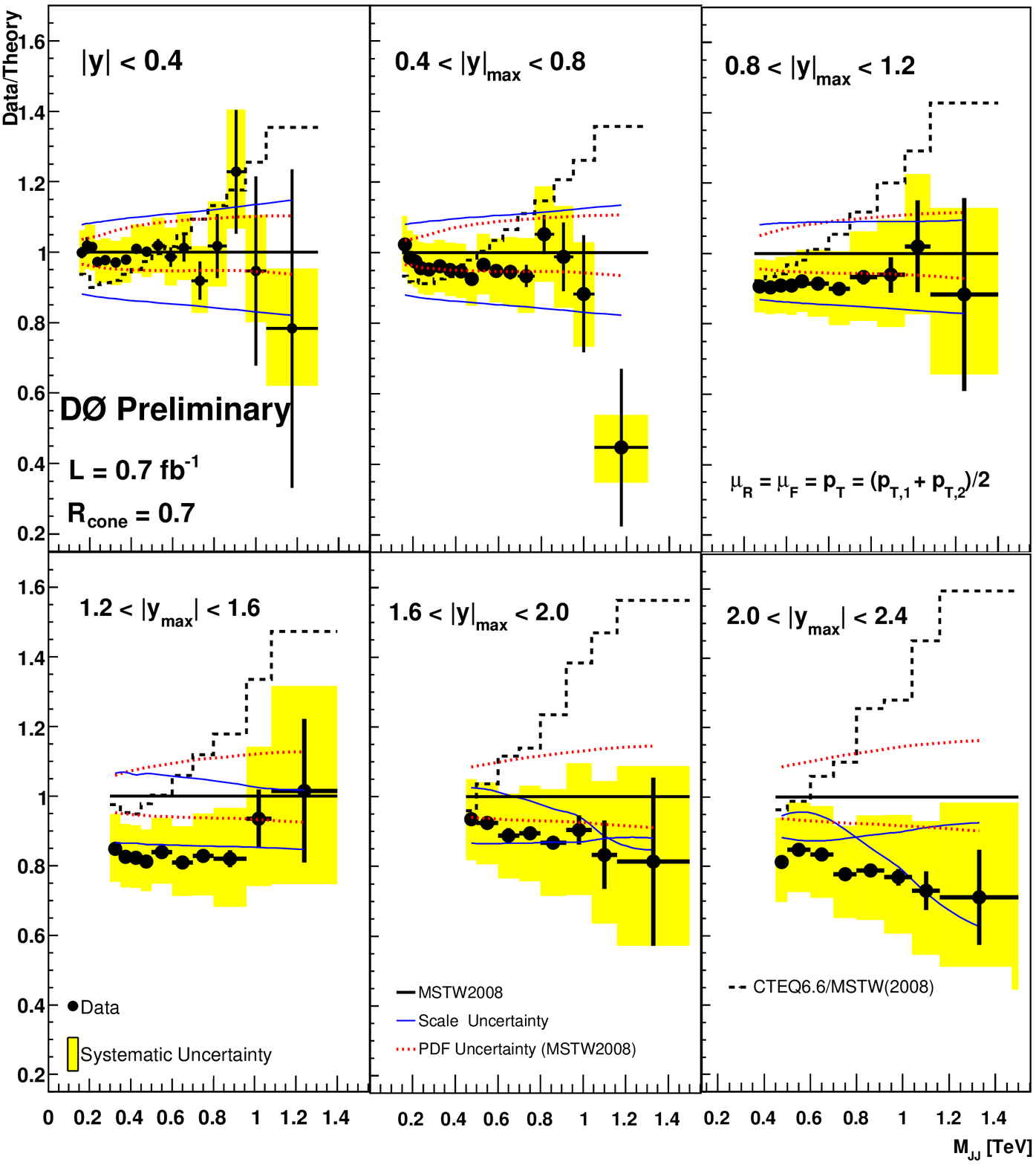}}
  \caption{{\em (top)} Measured dijet mass differential cross sections by CDF
    compared to NLO QCD predictions.
    {\em (bottom)} Ratios of the measured cross sections
    over the NLO QCD predictions from D0.
    \label{fig:mjj}}
\end{figure}

Both CDF~\cite{CDFMjjPRDRC} and D0~\cite{D0Dijets} also made
measurements of dijet mass differential cross sections.
The CDF measurement is based on 1.13~$\mbox{fb}^{-1}$ of
data~\cite{CDFMjjPRDRC} and jets are in the rapidity region
of $|y_{jet1,2}|<1$.
The measured dijet mass spectrum was found to be in good agreement
with NLO QCD predictions 
within the uncertainties as shown in Fig.~\ref{fig:mjj}.
Limits at the 95\% confidence level 
on the cross sections for new particles decaying into two jets
have been obtained based on this measurement~\cite{CDFMjjPRDRC}.
The mass exclusions are obtained for
the excited quarks, color-octet $\rho_T$, $W'$, $Z'$,
axigluons, and flavor-universal colorons.
For all strongly-produced particles, this analysis
yields the most stringent mass exclusions to date.

The D0's measurement extends to the more forward (large $|y|$)
region up to $2.0<|y_{max}|<2.4$ with the reduced jet energy
scale uncertainty~\cite{D0Dijets}. $|y_{max}|$ refers to the 
largest $|y|$ of the leading two jets.
In the forward region, the possible new physics contributions
are expected to be much reduced compared to the central (small
$|y|$) region, thus this extension makes tests of QCD
more stringent.
As shown in Fig.~\ref{fig:mjj}, 
the D0 measurement favors a new PDF set, MSTW2008~\cite{MSTW2008},
compared to CTEQ6.6~\cite{CTEQ66} 
which hasn't incorporated the inclusive jet cross section
measurements from the Tevatron Run II discussed above.

\begin{figure}[tbh]
\centerline{\includegraphics[width=0.70\columnwidth]{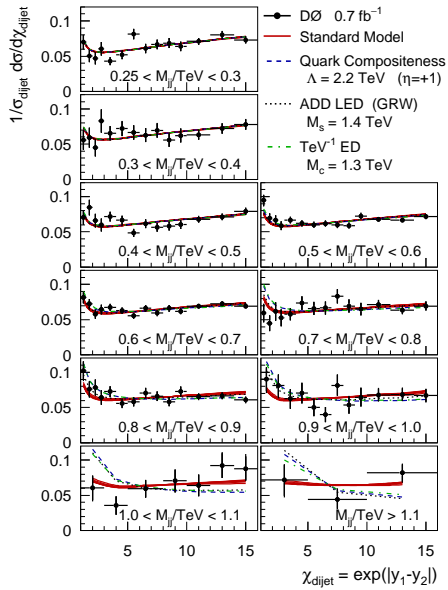}}
\caption{Ratios of the measured dijet angular distributions
  over NLO QCD predictions.}\label{fig:chi}
\end{figure}

CDF~\cite{CDFChi} and D0~\cite{D0Chi} also made measurements on
the dijet angular distribution.
D0 measurement covers a range in dijet mass from 0.25 TeV
to above 1.1 TeV, thus this is the first measurement
of angular distribution of a hard partonic scattering above
1 TeV in collider experiments.
The measurement is found to be in good agreement
with perturbative QCD predictions as shown
in Fig.~\ref{fig:chi}, and is used to
constrain new physics models including quark compositeness,
large extra dimensions, and TeV$^{-1}$ scale extra dimensions.
For all these models, this analysis
sets the most stringent direct limits to date.

\section{Inclusive $\gamma$, $\gamma$+jets, and $\gamma$+heavy flavor}
\label{sec:photon_jets}

The direct photon is a powerful probe of a hard scattering, as
it is not affected by parton fragmentation and hadronization.
Direct photon production (in association with jets)
is studied extensively by both HERA and Tevatron experiments.
Photons can be produced promptly in a hard scattering 
and they can be also produced in quark fragmentation.
In experimental measurements, such fragmentation contributions
are strongly suppressed by selecting ``isolated'' photons
which do not have other particles in their vicinities.

The prompt photon production was measured by H1 recently in
photoproduction~\cite{Collaboration:2009uj}, and the
measurements are compared to 
a QCD calculation based on the collinear
factorization in NLO (FGH)~\cite{Fontannaz:2003yn} and to a calculation based on the $k_T$
factorization approach (LZ)~\cite{Lipatov:2005tz}.
Neither of the calculations describes the data satisfactorily;
both calculations are below the data, most significantly
at low $E_T$.
The LZ calculation gives a reasonable description of the shape
of $\eta^{\gamma}$, whereas the FGH calculation is
significantly below the data for central and backward photons
($\eta^{\gamma} < 0.9$),
indicating the limitations of the current theoretical models.
More direct photon studies by HERA experiments were presented
by J. Ferrando~\cite{JamesFerrando}.

\begin{figure}[tp]\centering\leavevmode
  \centerline{
    \includegraphics[width=0.500\hsize,bb= 0 0 544 700,clip=]
    {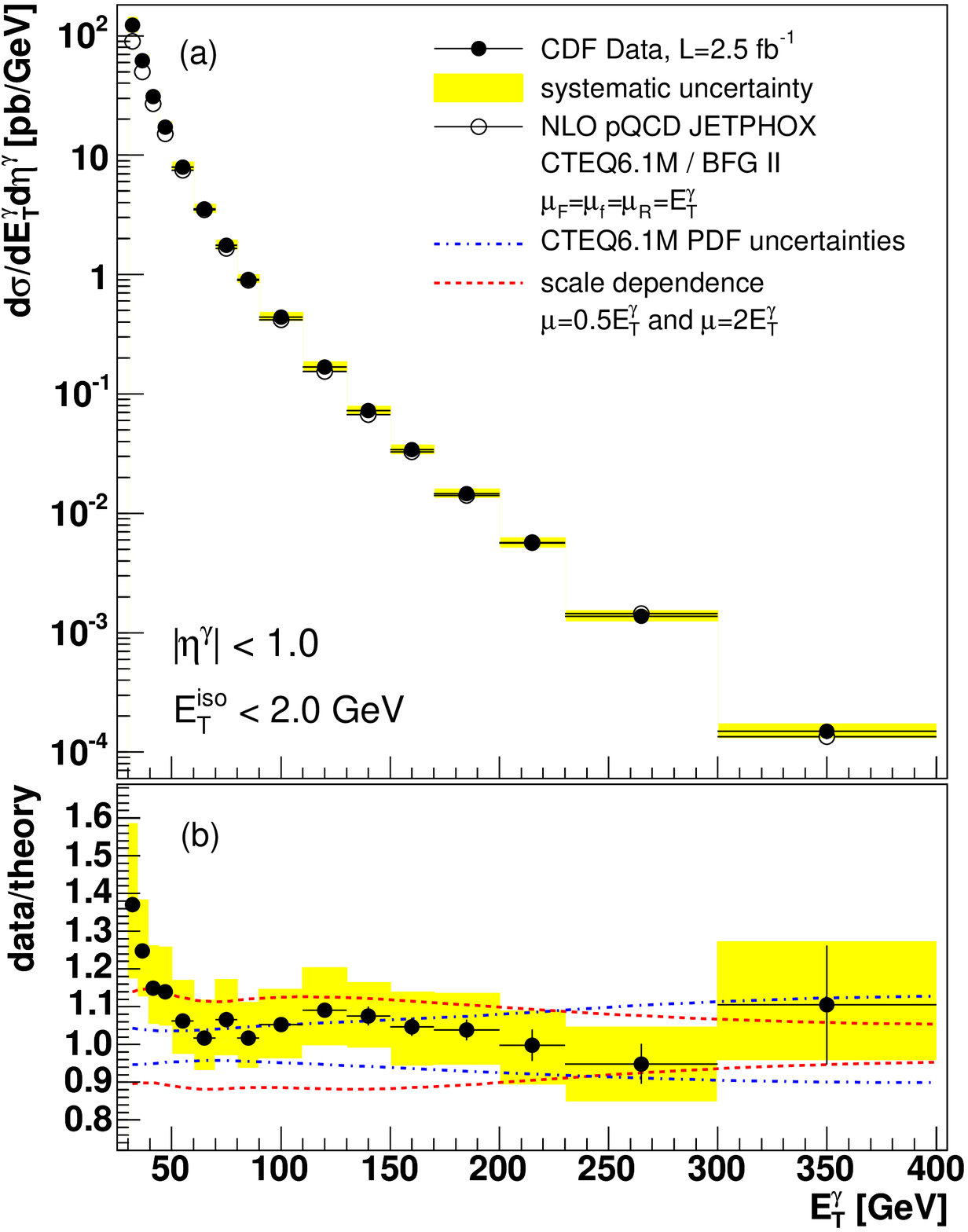}
    \includegraphics[width=0.490\hsize,bb= 25 -50 515 460,clip=]
    {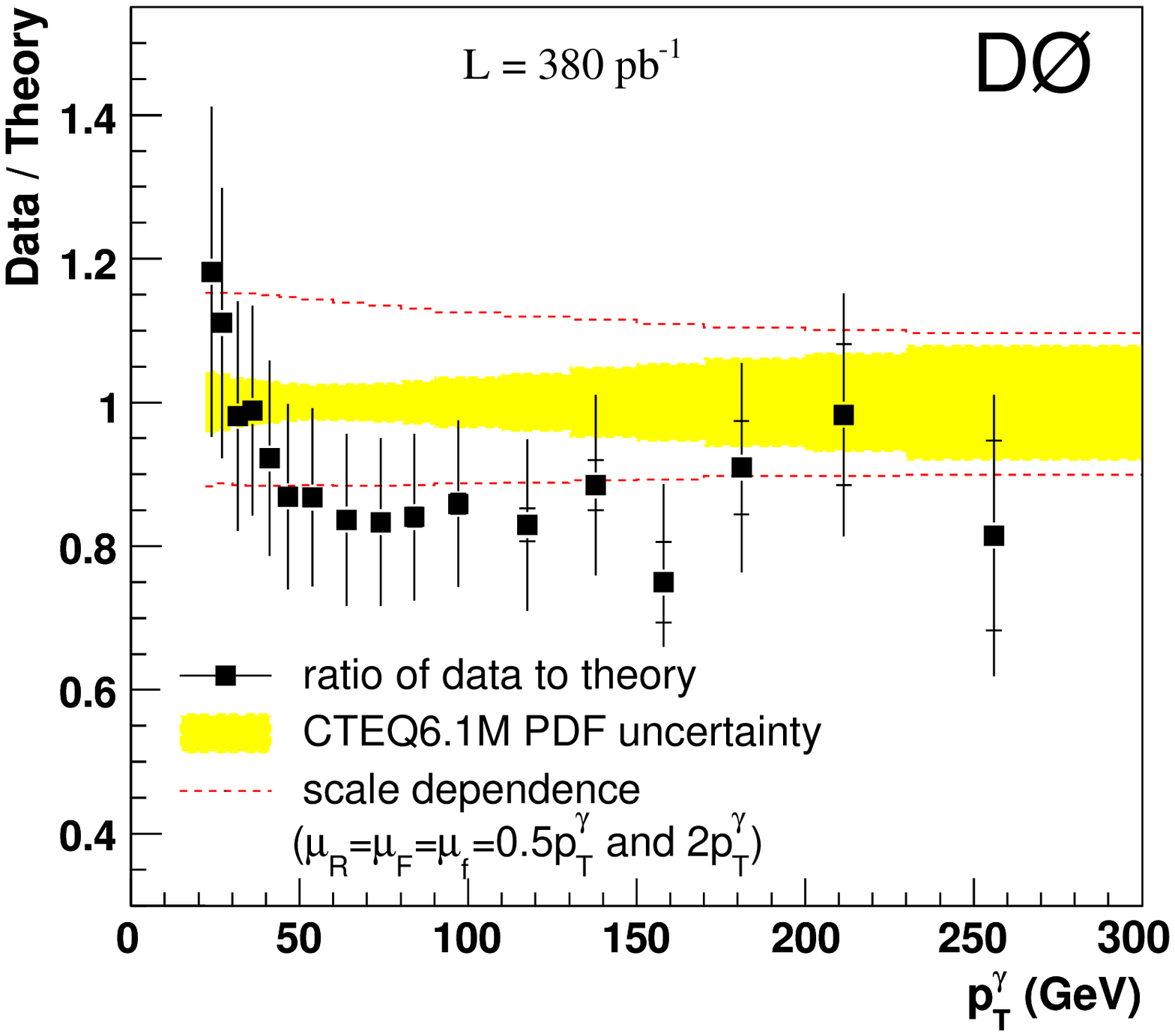}
    }
  \caption{The inclusive photon cross section measured by CDF and D0
    compared to NLO QCD predictions.}
  \label{fig:incl_photon}
\end{figure}

The inclusive photon cross section was also measured by
CDF~\cite{incl_pho_CDF} and D0~\cite{incl_pho_D0} 
at the Tevatron.
The direct photons predominantly come from
$\gamma$+jet production via Compton scattering
$q+g\to q+\gamma$ at low $p_T$ and
$q\bar q \to g\gamma$ at high $p_T$.
As shown in Fig.~\ref{fig:incl_photon},
the measured differential cross sections by both
CDF and D0 agree with NLO QCD calculations
within uncertainties at photon $E_T\gtrsim50$ GeV;
however, there is a rise at low $E_T$ in both
measurements and the similar trend has been
observed in earlier measurements at collider
and fixed target experiments.
This still needs to be understood theoretically.

To get further insight, D0 
investigated $\gamma$+jet event properties~\cite{gammajet_D0}.
As mentioned above, events in the inclusive photon cross section
measurements are mostly form $\gamma$+jet events; however,
in this measurement, 
the more complete hard scattering kinematics are investigated.
Measurements are made for the cases in which the photon
and jet are in the opposite side and same side in rapidity,
and jets are in the central region $|y_{jet}|<0.8$ and forward
$1.5<|y_{jet}|<2.5$ respectively; each topology is sensitive
to different parton $x$ values.
It was found that the data are not well described by
NLO QCD calculations, and the measurements will help
improve the theoretical descriptions of direct photon
production.

The $\gamma+b/c$ production is another interesting process.
Similarly to the non-heavy-flavor case,
$\gamma+b/c$ events are
produced predominantly via Compton scattering
$Q+g\to Q+\gamma$ (where $Q=b/c$) at low $p_T$,
and the contribution of the annihilation process $q\bar q
\to g\gamma \to Q\bar Q \gamma$ increases with increasing $p_T$.
The cross section is sensitive to the $c/b$-quark density
in the proton
which is so far indirectly extracted from gluon density.
The good understanding of this process is also important as
the QCD production of $\gamma+b/c$ is a significant background
in new physics searches, including searches for
$\omega_T$ production
($\omega_{T}\to\gamma\pi_{T}\to \gamma b\bar b$),
some SUSY scenarios, and excited $b$-quark production.

The $\gamma+b$-jet cross section was measured by
CDF~\cite{gammab_CDF} and D0 reported measurements on
$\gamma+b/c$-jet recently~\cite{gammaHF_D0}.
The measurement is made in the kinematic region of
$30<p_T^{\gamma}<150$ GeV,
$|y^{\gamma}|<1$,
$p_T^{jet}>15$ GeV, and
$|y^{jet}|<0.8$
for events with $y^{\gamma}y^{jet}>0$ and
$y^{\gamma}y^{jet}<0$, separately.
These two rapidity combinations help to differentiate the parton $x$
regions contributing to the two combinations. 

\begin{figure}[tp]\centering\leavevmode
  \includegraphics[width=0.740\hsize]
  {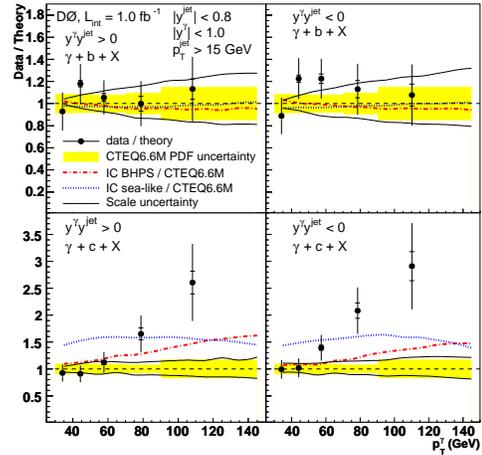}
  \caption{The $\gamma+b$-jet {\em (top)} and $\gamma+c$-jet {\em (bottom)}
    cross section
    ratio of data to theory as a function of $p_T^{\gamma}$ from D0.}
  \label{fig:photonHF}
\end{figure}

The measured differential cross sections for the $\gamma+b/c$-jet
production are compared to their theoretical predictions from NLO QCD
in Fig.~\ref{fig:photonHF}.
For $\gamma+b$, data and theoretical predictions are in good agreement
over the full kinematic region explored; however, 
for $\gamma+c$, a reasonable agreement is observed only at
$p_T^{\gamma}<70$ GeV, and the deviation increases with increasing
$p_T^{\gamma}$ in both event topologies.
The deviation may be attributed to a possible non-negligible intrinsic
charm content in the proton and/or the inaccurate description of
$g\to c\bar c$ fragmentation.

\section{W/Z+jets and W/Z+heavy flavor}
\label{sec:wz_jets}

As discussed in Sec.~\ref{sec:intro},
event samples of $W/Z$+jets are important for Tevatron and LHC
physics, and good understanding of these processes via QCD
influences many precision measurements and the Higgs and BSM
discovery potentials.
The perturbative QCD calculations and many MC models are
available for $W/Z$+jets production.
The experimental measurements have been made to validate them
or to help improve the models.

\begin{figure}[th]\centering\leavevmode
  \begin{tabular}{c}
  \includegraphics[width=0.60\hsize,clip=]
  {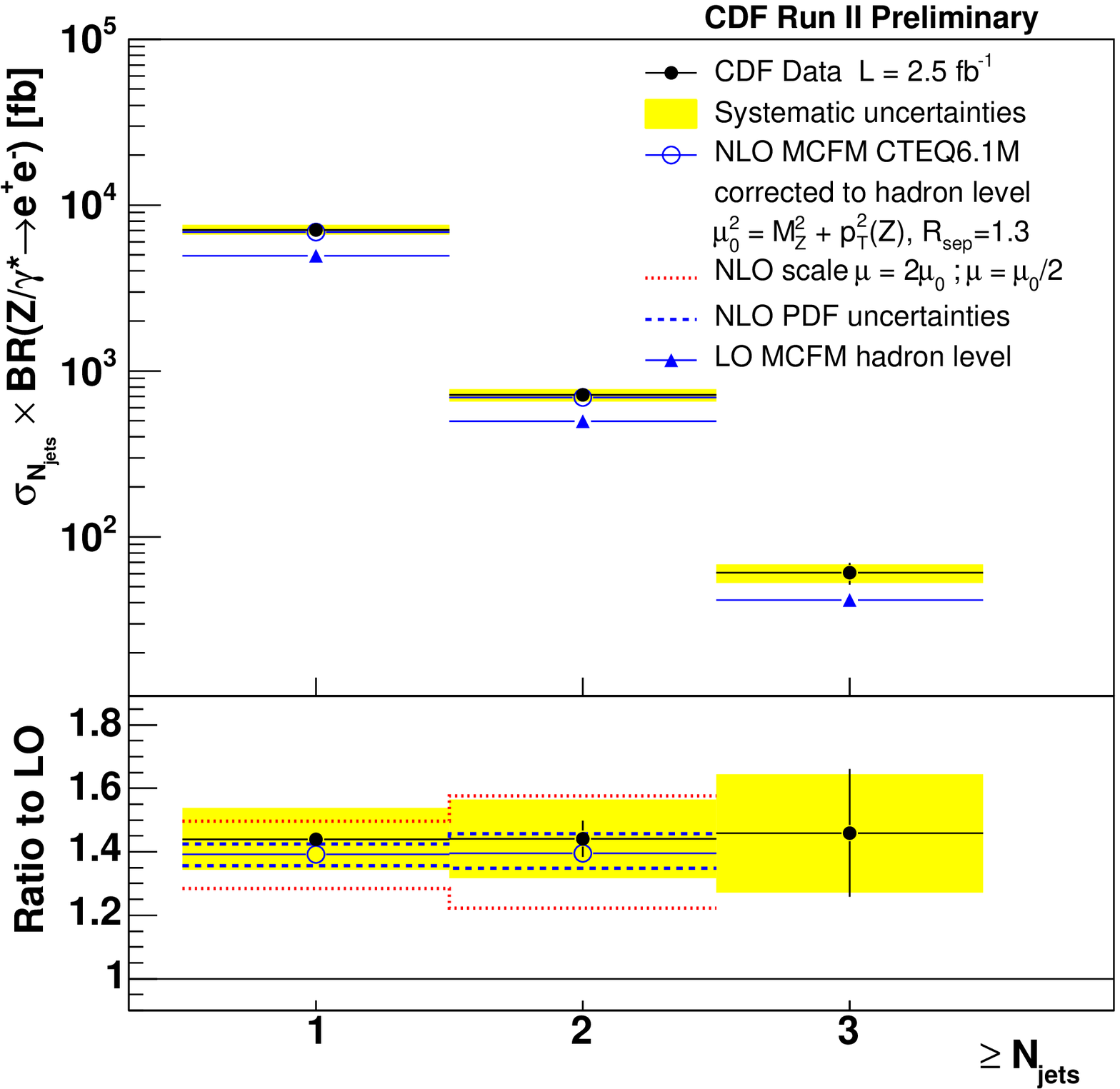}\\
  \includegraphics[width=1.00\hsize,clip=]
  {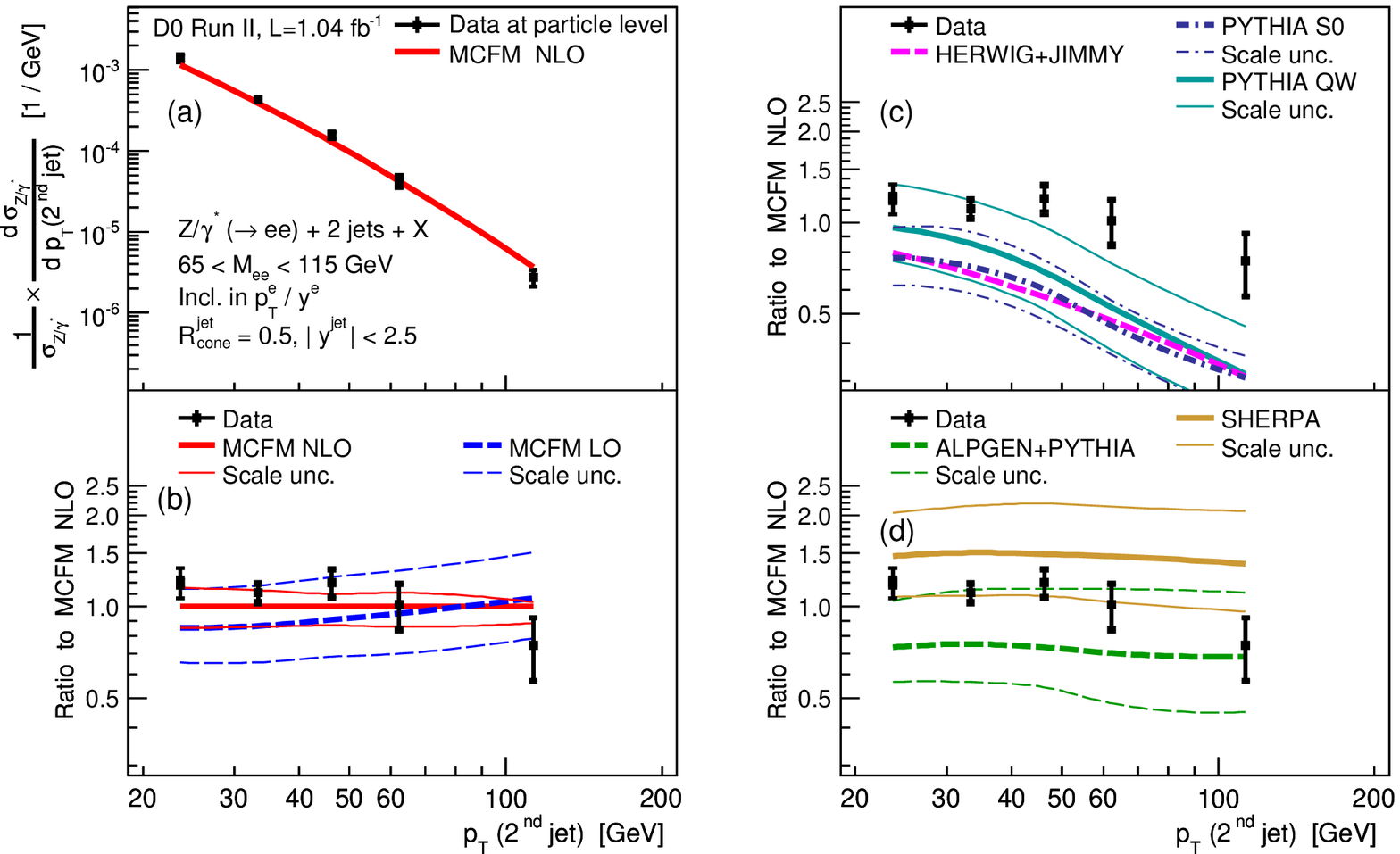}
  \end{tabular}
  \caption{
    {\em (top)} The measured $Z(\to ee)$+jets cross section versus the
    inclusive jet multiplicities from CDF.
    {\em (bottom)} The measured $Z(\to ee)$+jets differential cross section
    for the second jet in $Z$+2 jets+$X$ events,
    compared to the predictions of {\sc mcfm} (a,b),
    three parton-shower event generator models (c), 
    and two event generators matching matrix-elements 
    to a parton shower (d).}
  \label{fig:Zjets}
\end{figure}

CDF made measurement on $W$+jets production~\cite{Wjets_CDF} up to
$\ge4$ jets, and later
CDF~\cite{Zjets_CDF} and D0~\cite{Zjets_D0} made measurement on $Z$+jets
production up to $\ge3$ jets as well.
In all these measurements, $W/Z$ are identified with their
leptonic decays ($W\to l\nu,~Z\to ll,~l=e\mbox{ and/or }\mu$).
In LO QCD,
Z+jet events are produced predominantly by the 
processes $gq\to Zq$ and 
$q\bar q\to Zq$, and additional parton radiation
may produce multiple jets.
As shown in Fig.~\ref{fig:Zjets},
data are consistent with NLO QCD predictions 
from {\sc mcfm} where they
are available {\it i.e.} up to the inclusive
multiplicity bin of $n_{jet}\ge2$.
The data also suggest that NLO/LO ratio is approximately
constant across the inclusive jet multiplicities which may 
mean that we can use LO models reliably
with a constant $K$-factor correction.

As shown in Fig.~\ref{fig:Zjets},
the D0 measurement~\cite{Zjets_D0} presents detailed
comparisons with NLO QCD and several popular MC event
generator models, {\sc alpgen}+{\sc pythia}, {\sc sherpa}, 
{\sc pythia} with S0~\cite{Skands:2007zg} and 
QW~\cite{Albrow:2006rt} tunes, and {\sc herwig}+{\sc jimmy}.
The parton-shower based {\sc herwig} and {\sc pythia}
generator models show significant disagreements
with data which increase with the jet multiplicity in events,
and, the {\sc sherpa} and {\sc alpgen}+{\sc pythia}
generators show an improved description
of data as compared with the parton-shower-based
generators. 
{\sc alpgen}+{\sc pythia} and {\sc sherpa} predict
lower and higher rates than observed in data, respectively;
however, by changing the $\mu_R$ and $\mu_F$ scales by a factor
of $\sim 2$ the agreement improves to an acceptable level.

The $W+c$ production has been studied by both
CDF~\cite{Wc_CDF} and D0~\cite{Wc_Dzero}, and 
CDF has also made a measurement of the $W+b$-jets cross
section~\cite{Wb_CDF} in the $W\to l\nu~(l=e,\mu)$ channel,
recently.
The cross section for $b$-jets with $E_T>20$ GeV and $|\eta|<2$
from QCD $W+b$ production is measured to be
$\sigma(Wb)\times BR(W\to l^{\pm}\nu)
=2.74 \pm 0.27\mbox{(stat)} \pm0.42\mbox{(syst)} \mbox{~pb}$
which is to be compared to 
the {\sc alpgen} prediction of 0.78 pb and {\sc pythia}
prediction of 1.10 pb;
showing the differences of factors of 2.5-3.5.
The NLO QCD prediction is $1.22\pm0.14$ pb
which is also low with respect to the measured value.
The upcoming differential measurement may tell us more
about physics behind these observed differences.

The $Z+b$-jets production has been another important process,
and it has been studied by both D0~\cite{Zb_Dzero}
and CDF~\cite{Zb_CDF}.
The dominant production diagrams contributing to the $Z+b$-jets
final state are (a) $bg\to Zb$ ($\sim65$\%) and
(b) $q\bar q\to Zb\bar b$ ($\sim35$\%) in
NLO QCD predictions.
The cross section is sensitive to the $b$-quark density.
A good understanding of the $b$ density is essential
to accurately predict the production of particles that couple strongly to
$b$-quarks including SUSY Higgs bosons.
$Z+b$-jets production is also a major background in searches for the
Higgs production in the $ZH\to Zb\bar b$ channel.

\begin{figure}[th]\centering\leavevmode
  \includegraphics[width=0.60\hsize,bb=1 280 568 561,clip=]
  {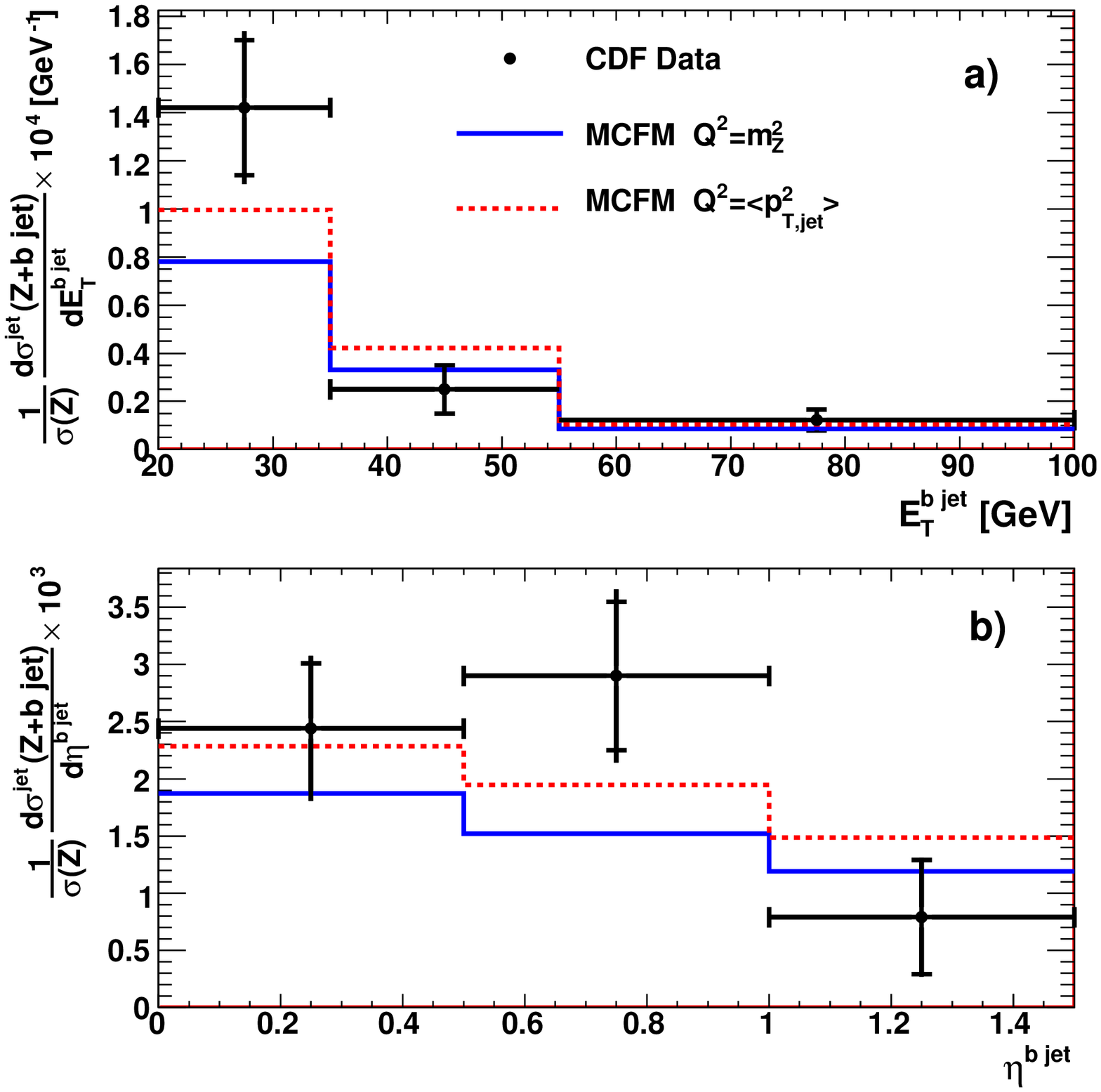}\\
  \includegraphics[width=0.60\hsize,bb=1 280 568 561,clip=]
  {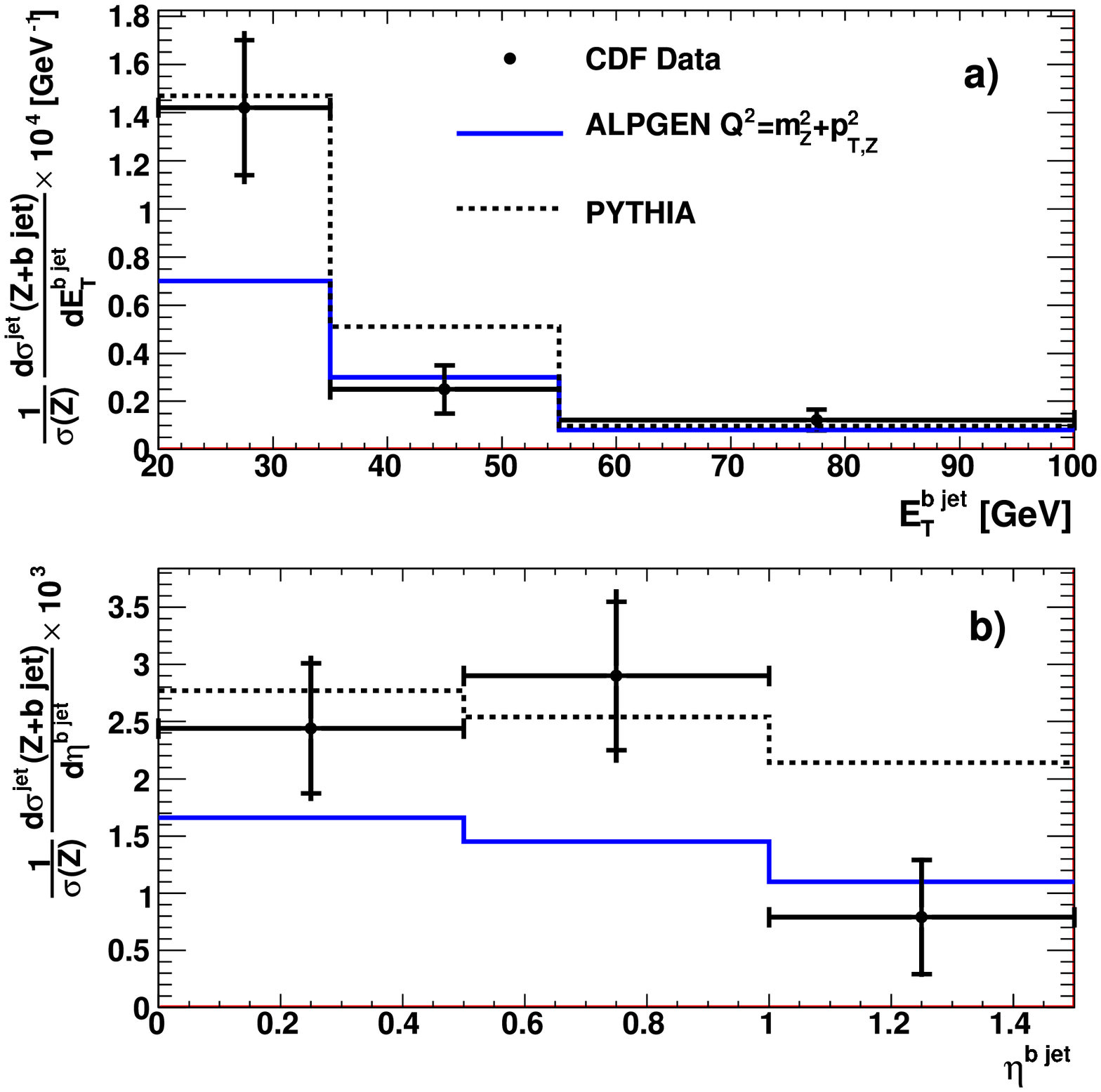}
  \caption{$Z+b$-jets differential cross sections as a function of jet
    $E_T$ measured by CDF compared to theoretical predictions.}
  \label{fig:Zb}
\end{figure}

Recent CDF measurement from 2.0 fb$^{-1}$ of data
was made using jets with $E_T>20$ GeV and $|\eta|<1.5$
in the $Z\to ll~(l=e,\mu)$ channel.
The measured cross section ratio is
$\sigma(Z+b)/\sigma(Z+\mbox{jets})=2.08 \pm 0.33 \pm 0.34$\%,
which should be compared to predictions of 
1.8\% from {\sc mcfm} with $Q^2 = m_Z^2+p_{T,Z}^2$, 
2.2\% with $Q^2 = \langle p_{T,jet}^2 \rangle$, 
1.5\% from {\sc alpgen}, 
and 2.2\% from {\sc pythia}.
The high statistics data used in the recent analysis
allowed the first measurement of differential distributions
in this process
which are shown in Fig.~\ref{fig:Zb} together with several
theoretical predictions.
All predictions are generally in agreement with the data, but
differences of up to 2$\sigma$
are observed in the integrated cross section between the data
and the {\sc mcfm} calculation, depending on which scale is used.
The large spread of the theoretical predictions suggests
that higher orders in QCD calculation may be important
for this process.


\section{Underlying Event and Multiple Parton Interactions}
\label{sec:ue}

%
%

Many of the important observables at hadron colliders,
including jets,
are sensitive to the underlying event (UE), so a good
understanding of the UE is needed for precision measurements
and for improving discovery potential.
The UE is considered to arise from the multiple parton
interactions in a hadron-hadron collision.
Many studies on the UE have been made by the
CDF~\cite{CDFUE} and H1~\cite{H1UE} experiments.

\begin{figure}[tbh]\centering\leavevmode
\begin{center}
\includegraphics[width=.50\columnwidth]{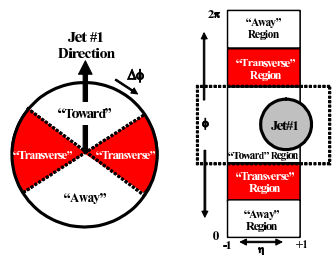}
\end{center}
\caption{Definitions of the toward, away and transverse regions.
The angle $\Delta\phi=\phi-\phi_{jet1}$ is the relative azimuthal
angle between particles and the direction of the leading jet or $Z$,
and each region is considered for $|\eta|<1$.}%
\label{fig:ue_ana_topology}%
\end{figure}

\begin{figure}[tbh]\centering\leavevmode
\centerline{
  \includegraphics[height=0.495\columnwidth,angle=270,bb=60 20 580 730]{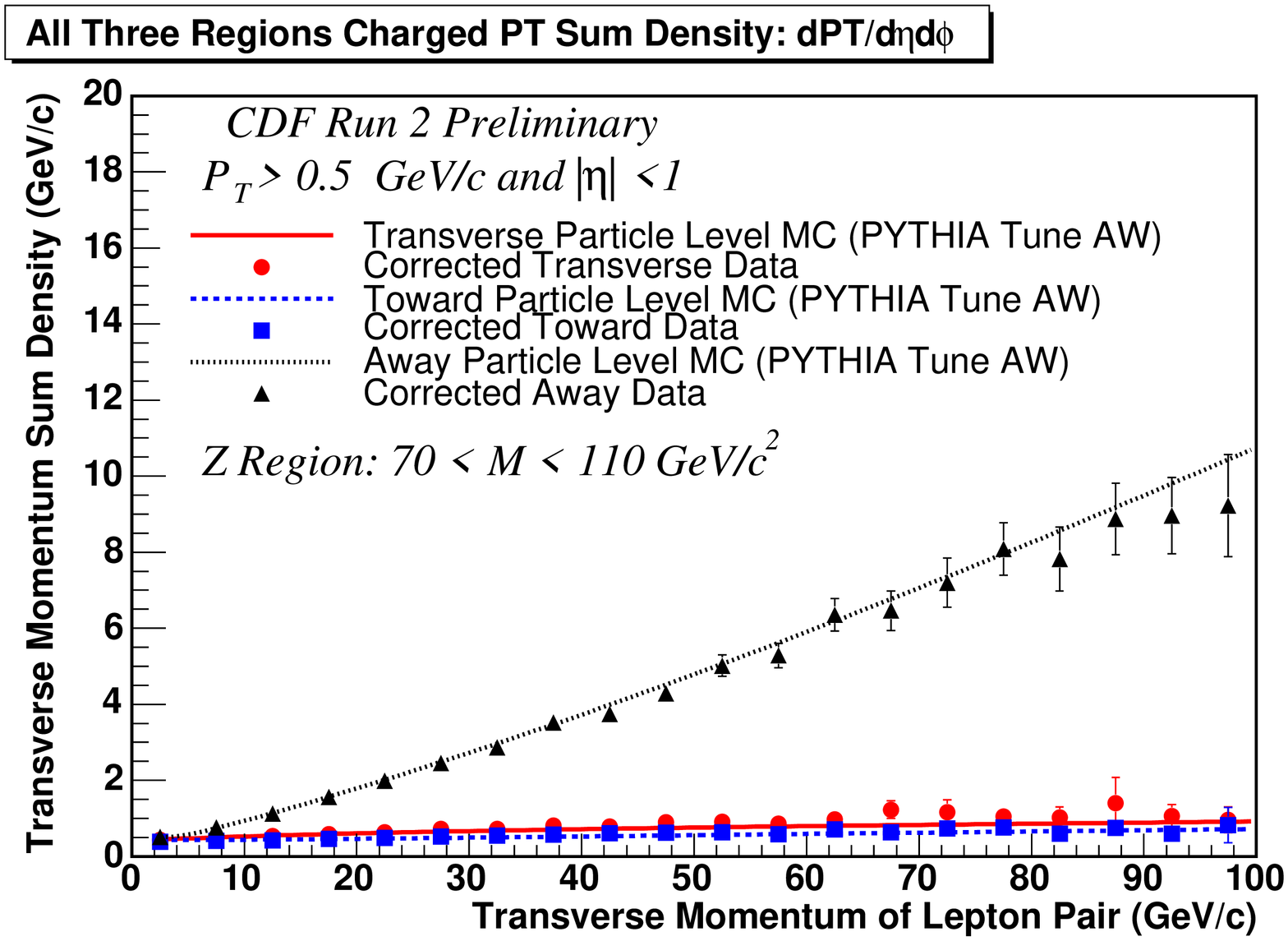}
  \includegraphics[height=0.495\columnwidth,angle=270,bb=60 20 580 730]{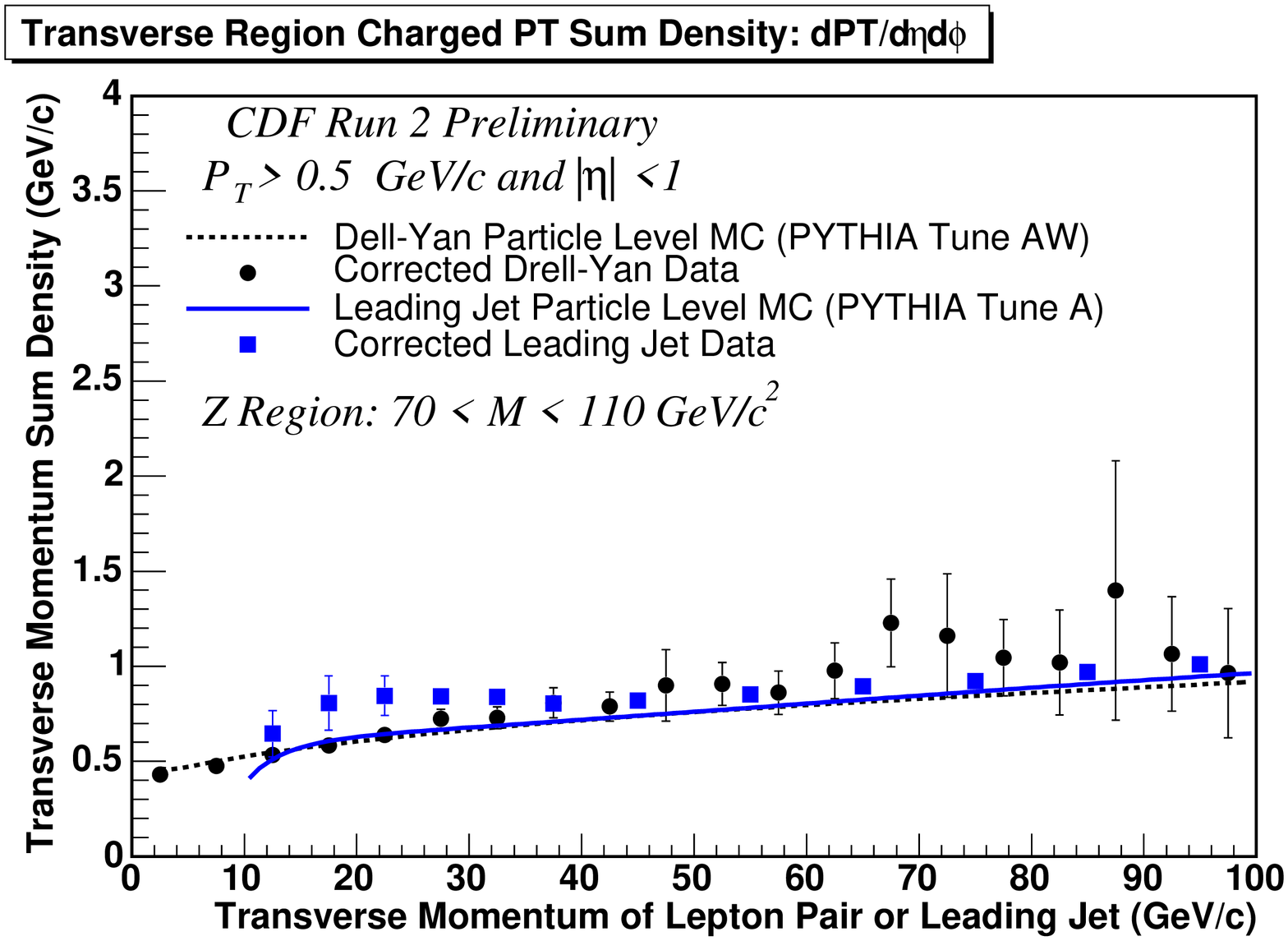}}
\caption{{\em (left)} $d^2p_T/d\eta d\phi$ in the three regions in Drell-Yan production.
{\em (right)} Comparison the transverse region $d^2p_T/d\eta d\phi$ between Drell-Yan and
jet production.\label{fig:ue}}
\end{figure}

CDF measurements in jet and Drell-Yan production
have made use of the topological structure of
hadron-hadron collisions to study the UE
as shown in Fig.~\ref{fig:ue_ana_topology}.
In jet events, measurements on event activities in the
transverse region with respect to the jet axis 
are most sensitive to the UE.
In Drell-Yan events, not only the transverse region but also the
toward region 
can be used to study the UE after excluding
two leptons from Drell-Yan production.
%
%
%
%

The charged particle $p_T$ density $d^2p_{T}/d\eta d\phi$ 
in the toward and transverse regions in Drell-Yan events
is shown as a function of the lepton pair $p_T$ in
Fig.~\ref{fig:ue}.
The Drell-Yan lepton pairs have the invariant mass around the
$Z$ mass ($70<m_{ll}<110$ GeV).
The distributions are found to be rather flat with the increasing
lepton pair $p_T$ in the toward and transverse regions,
but goes up in the away region to balance lepton pairs as expected,
and these are well described by recent tuned {\sc pythia}~\cite{tunes}.

In Fig.~\ref{fig:ue}, $d^2p_{T}/d\eta d\phi$ in the transverse
region in Drell-Yan and jet events are compared as a function of
lepton pair or leading jet $p_T$.
At low $p_T$, Drell-Yan events still have the large invariant mass
lepton-pairs,
whereas we only get soft interaction events in a jet event sample.
So, dijet and Drell-Yan events have distinct topologies; nevertheless,
the two distributions show a similar trend, indicating the universality
of the UE.

In these measurements, there are many more distributions
of observables sensitive to the UE corrected to the particle
level, and will provide important knobs to tune MC models.
A recent measurement of inclusive $p\bar p$ interactions 
by CDF~\cite{CDFMB} also provides important distributions for MC tuning.
Especially, the correlation between the charged particle
multiplicity and average $p_T$ has been found to be sensitive
to subtle changes of {\sc pythia} tuning parameters and have been
popularly used phenomenologists~\cite{Skands:2007zg}.

In the meantime, D0 has made a new precise measurement on 
double parton scattering, {\it i.e.} hard interactions of
two parton pairs in a single $p\bar p$ interaction.
This measurement provides insight to parton spatial
distributions in the proton, and this can be a non-negligible
background to other processes especially at high instantaneous
luminosities.
The measurement is made using $\gamma$+3 jets events, and
the cross section is defined as $\sigma_{\rm DP}
=\sigma_{\gamma j} \sigma_{jj} / \sigma_{\rm eff}$,
where $\sigma_{\rm eff}$ is the effective interaction region.
The larger $\sigma_{\rm eff}$ means partons are more uniformly
distributed.
The $\sigma_{\rm eff}$ is measured to be $15.1\pm1.9$ mb
which is consistent with earlier measurements 
and does not show dependence on jet $p_T$ from the second
interaction.

\section{Expected Results from LHC} 
\label{sec:lhc}

As a variety of new results are arriving from HERA
and Tevatron, the Large Hadron Collider at CERN, Geneva,
Switzerland, is expected to start its operation
in coming months and the LHC experiments will start
delivering physics results.
The significant increase in the hadron-hadron collision
energy from the Tevatron to the LHC enhances physics
potentials; potentials for precision measurements and
the discovery potentials for the Higgs boson and
various BSM models.

One of the early physics measurements which will be made at the
LHC experiments is on inclusive $pp$ inelastic collisions and 
underlying event which will build basis for
high $p_T$ physics program.
The extrapolation from the lower $\sqrt{s}$
experiments to the LHC has the sizable uncertainty,
and this needs 
more precise measurements.

\begin{figure}[tp]\centering\leavevmode
  \centerline{
    \includegraphics[width=0.70\columnwidth,bb=66 517 300 734,clip=]
    {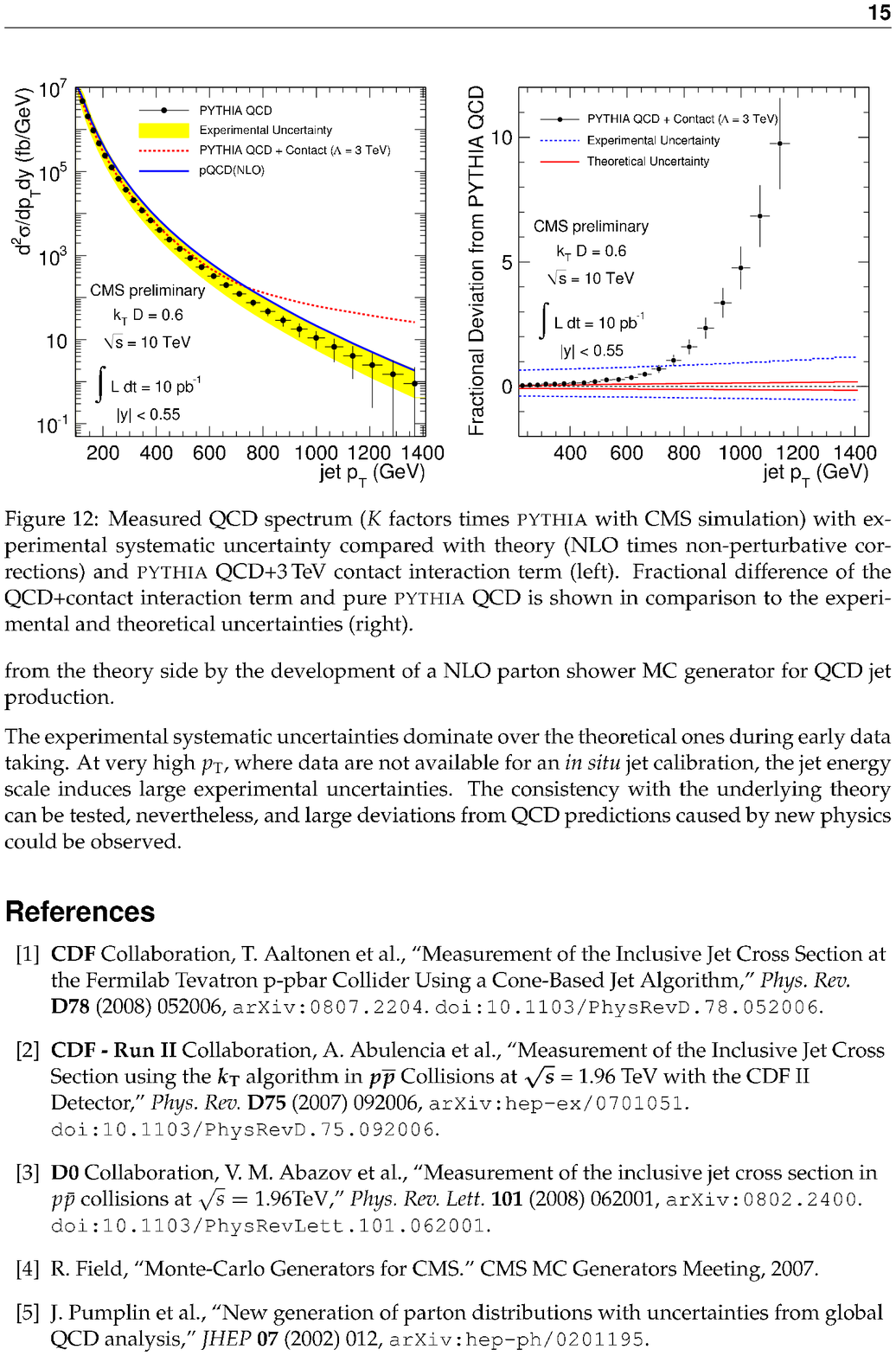}
  }
\caption{Inclusive jet cross differential cross section
expected to be measured in $\sqrt{s}=10$ TeV $pp$ collision data
along with its experimental systematic uncertainty.
It is compared with NLO QCD predictions and {\sc pythia} QCD+3 TeV
contact interaction term.}
\label{fig:CMS-IncJet}
\end{figure}

Various jet physics measurements are also planed
and an expected result on the inclusive jet cross section
measurement from the CMS experiment is shown in 
Fig.~\ref{fig:CMS-IncJet}.
On day 1, CMS assumes 10\% jet energy scale uncertainty
which results in a uncertainty of a factor $\sim2$ in
differential cross section.
However, still the measurement has a good chance to probe quark
compositeness beyond the Tevatron reach with 10 $\mbox{pb}^{-1}$.
In the absence of BSM contributions, the
measurement will constrain PDFs and lead to
$\alpha_s$ extraction at high $Q^2$.

This is one example from jet physics studies at the LHC, and
many more results will arrive from the LHC experiments 
in coming years.

\section{Summary and Remarks} 

Tremendous effort has been made to advance understanding of QCD
in studies of high $E_T$ jet events and also events with jets
plus $\gamma$, $W$, $Z$ from experiments at the HERA $ep$ collider
and Tevatron $p\bar p$ collider.

The measurements of jet cross sections at HERA and Tevatron
enable us to extract precisely the strong coupling constant,
$\alpha_s$, the most fundamental parameter in QCD.
They also provide valuable constraints on parton distribution
functions (PDFs) of the proton.
Some of them put constraints on BSM physics models as well.

Among the measurements including photons,
$\gamma$+jets and $\gamma$+$c$-jets measurements show
disagreements from NLO QCD predictions
beyond the experimental and estimated theoretical
uncertainties.
These need to be resolved before we take advantage of
these measurements for determination of the proton PDFs.

Accurate modeling of $W$/$Z$+jets event topology in hadron-hadron
collisions, especially at high jet multiplicity or when jets are
originated from heavy flavor, is challenging theoretically;
however it is also critical since such events are very important
backgrounds to many physics analyses.
%
%
The measurements with improved precision from the Tevatron experiments
would be the key for deeper understanding of these processes and
benefit for future physics analyses at the Tevatron and also at the
LHC.

In this paper, measurements sensitive to the UE,
components in the hadron collision event not associated with the
hard scattering, from HERA and Tevatron are also discussed.
They play an important role in obtaining an accurate model of
the underlying event, and such accurate model is critical
in jet production and many other high-$p_T$ hadron
collider physics studies.

While HERA finished its operations in 2007, the
Tevatron is expected to continue its operation through 2011
and deliver up to 12 $\mbox{fb}^{-1}$ of data which is
more than 5 times the one used in the studies presented
in this paper. Much more interesting results are expected
to arrive from these experiments.

In the meantime, we also expect first results from the LHC
as early as the next year.
The $pp$ center-of-mass energy of up to
$\sqrt{s}=14$ TeV at the LHC will allow testing the SM at
unprecedented high $Q^2$ and have good potential to
discover physics beyond the SM.
The physics knowledge and analysis developments made
at HERA and Tevatron have enhanced the LHC physics
potentials. 
Let's stay tuned for new results in coming
years.

\section{Acknowledgment} 

I would like to thank C. Mesropian, S. Pronko, A. Bhatti,
J. Dittmann, M. Wobisch, D. Lincoln, G. Hesketh, C. Glasman,
G. Grindhammer, J. Ferrando, S. Tapprogge for their helpful
inputs in preparation of this paper.



\end{document}